\documentclass[preprint,pre,aps,showpacs,amsmath]{revtex4}
\usepackage{graphicx}

\begin{document}

\title{Strange Nonchaotic Oscillations in the Quasiperiodically Forced Hodgkin-Huxley Neuron}

\author{Woochang Lim $^{a,}$}
\email{wclim@kangwon.ac.kr}
\author{Sang-Yoon Kim $^{a,b,}$}
\thanks{Corresponding Author}
\email{sykim@kangwon.ac.kr} \affiliation{$^{a}$ Department of
Physics, Kangwon National University, Chunchon, Kangwon-Do
200-701, Korea\\ $^{b}$ Department of Physics, University of
Wisconsin-Milwaukee, P.O. Box 413, Wisconsin 53211, U.S.A.}

\begin{abstract}
We numerically study dynamical behaviors of the quasiperiodically
forced Hodgkin-Huxley neuron and compare the dynamical responses
with those for the case of periodic stimulus. In the periodically
forced case, a transition from a periodic to a chaotic oscillation
was found to occur via period doublings in previous numerical and
experimental works. We investigate the effect of the quasiperiodic
forcing on this period-doubling route to chaotic oscillation. In
contrast to the case of periodic forcing, new type of strange
nonchaotic (SN) oscillating states (that are geometrically strange
but have no positive Lyapunov exponents) are found to exist
between the regular and chaotic oscillating states as intermediate
ones. Their strange fractal geometry leads to aperiodic
``complex'' spikings. Various dynamical routes to SN oscillations
are identified, as in the quasiperiodically forced logistic map.
These SN spikings are expected to be observed in experiments of
the quasiperiodically forced squid giant axon.
\end{abstract}

\maketitle

\section{Introduction}
\label{sec:INT}

To probe dynamical properties of a system, one often applies an
external stimulus to the system and investigate its response.
Particularly, periodically stimulated biological oscillators have
attracted much attention in various systems such as the embryonic
chick heart-cell aggregates \cite{Chicken} and the squid giant
axon \cite{Squid,Aih}. These periodically forced systems have been
found to exhibit rich regular and chaotic behaviors \cite{Glass}.
Recently, similar lockings and chaotic responses have also been
found in neocortical networks of pyramidal neurons under periodic
synaptic input \cite{Stoop}. On the contrary, quasiperiodically
forced case has received little attention \cite{QO}. Hence,
intensive investigation of quasiperiodically forced biological
oscillators is necessary for understanding their dynamical
responses under the quasiperiodic stimulus.

Strange nonchaotic (SN) attractors typically appear between the
regular and chaotic attractors in quasiperiodically forced
dynamical systems
\cite{SNA,PNR,Greb,NK,HH,PD,PMR,Kim,Kim2,Kim3,Kim4,Exp}. They
exhibit some properties of regular as well as chaotic attractors.
Like regular attractors, their dynamics is nonchaotic in the sense
that they do not have a positive Lyapunov exponent; like usual
chaotic attractors, they have a geometrically strange fractal
structure. Here, we are interested in dynamical responses of
neural oscillators subject to quasiperiodic stimulation. SN
oscillations are expected to occur in quasiperiodically forced
neural oscillators.

This paper is organized as follows. In Sec.~\ref{sec:SN}, we study
dynamical responses of the quasiperiodically forced Hodgkin-Huxley
(HH) neuron which was originally introduced to describe the
behavior of the squid giant axon \cite{HH1}. As a dc stimulus
passes a threshold value, a self-sustained oscillation
(corresponding to a spiking state) is induced. Effect of periodic
forcing on this HH oscillator was previously studied
\cite{Aih,HH2}. Thus, regular (such as phase locking and
quasiperiodicity) and chaotic responses were found. We note that
similar dynamical responses were also observed in experimental
works of the periodically forced squid giant axon
\cite{Aih,Squid}. In this way, the model study on the HH neuron
may be examined in real experiments of the squid giant axon. Here,
we numerically study the case that the HH oscillator is
quasiperiodically forced at two incommensurate frequencies and
compare the dynamical responses with those for the periodically
forced case. In the periodically forced case ({\it i.e.}, in the
presence of only one ac stimulus source), a transition from a
periodic to a chaotic oscillation was found to occur via
period-doubling bifurcations \cite{Aih,HH2}. Effect of the
quasiperiodic forcing on this period-doubling route to chaotic
oscillation is particularly investigated by adding another
independent ac stimulus source. Thus, unlike the case of periodic
forcing, new type of SN oscillating states are found to occur
between the regular and chaotic oscillating states as intermediate
ones. As a result of their strange geometry, SN oscillating states
give rise to the appearance of aperiodic complex spikings, as in
the case of chaotic oscillations. Diverse dynamical routes to SN
oscillations are identified, like the case of the
quasiperiodically forced logistic map \cite{SNA}. It is also
expected that these SN spikings might be observed in experimental
works on the quasiperiodically forced squid giant axon. Finally, a
summary is given in Sec.~\ref{sec:SUM}.

\section{SN Oscillations in the Quasiperiodically Forced HH Oscillator}
\label{sec:SN}

We consider the conductance-based HH neuron model which serves as
a canonical model for tonically spiking neurons. The dynamics of
the HH neuron, which is quasiperiodically forced at two
incommensurate frequencies $f_1$ and $f_2$, is governed by the
following set of differential equations:
\begin{subequations}
\begin{eqnarray}
C\frac{dV}{dt} &=& -I_{ion} + I_{ext} =
-(I_{Na} + I_K + I_L) + I_{ext} \nonumber \\
&=& -g_{Na}m^3 h (V-V_{Na}) - g_K n^4 (V-V_K) -
g_L (V-V_L) + I_{ext}, \label{eq:HHa} \\
\frac{dx}{dt} &=& \alpha_x (V) (1-x) - \beta_x (V) x; \;\;\; x=m,
h, n, \label{eq:HHb}
\end{eqnarray}
\label{eq:HH1}
\end{subequations}
where the external stimulus current density (measured in units of
$\mu {\rm A /cm^2}$) is given by $I_{ext} = I_{dc} + A_1 \sin(2
\pi f_1 t) + A_2 \sin(2 \pi f_2 t)$, $I_{dc}$ is a dc stimulus,
$A_1$ and $A_2$ are amplitudes of quasiperiodic forcing, and
$\omega (\equiv f_2 / f_1)$ is irrational ($f_1$ and $f_2$:
measured in units of kHz). Here, the state of the HH neuron at a
time $t$ (measured in units of ms) is characterized by four
variables: the membrane potential $V$ (measured in units of mV),
the activation (inactivation) gate variable $m\,(h)$ of the $Na^+$
channel [{\it i.e.,} the fraction of sodium channels with open
activation (inactivation) gates], and the activation gate variable
$n$ of the $K^+$ channel ({\it i.e.,} the fraction of potassium
channels with open activation gates). In Eq.~(\ref{eq:HHa}), $C$
represents the membrane capacitance per surface unit (measured in
units of $\mu {\rm F/cm^2}$) and the total ionic current $I_{ion}$
consists of the sodium current $I_{Na}$, the potassium current
$I_K$, and the leakage current $I_L$. Each ionic current obeys
Ohm's law. The constants $g_{Na}$, $g_{K}$, and $g_L$ are the
maximum conductances for the ion and the leakage channels, and the
constants $V_{Na}$, $V_K$, and $V_L$ are the reversal potentials
at which each current is balanced by the ionic concentration
difference across the membrane. The three gate variables obey the
first-order kinetics of Eq.~(\ref{eq:HHb}). The rate constants are
given by
\begin{subequations}
\begin{eqnarray}
\alpha_m(V) &=& \frac{0.1[25-(V-V_r)]}{\exp [\{25-(V-V_r)\}/10] -
1}, \;\; \beta_m(V) = 4 \exp [-(V-V_r)/18], \\
\alpha_h(V) &=& 0.07 \exp [-(V-V_r)/20], \;\; \beta_h(V) =
\frac{1}{\exp [\{30-(V-V_r)\}/10]+1}, \\
\alpha_n(V) &=& \frac{0.01[10-(V-V_r)]}{\exp
[\{10-(V-V_r)\}/10]-1}, \;\; \beta_n(V) = 0.125 \exp
[-(V-V_r)/80],
\end{eqnarray}
\label{eq:RC}
\end{subequations}
where $V_r$ is the resting potential when $I_{ext}=0$. For the
squid giant axon, typical values of the parameters (at 6.3
${}^\circ{\rm C}$) are \cite{HH3}: $C=1 \mu{\rm F/cm^2}$,
$g_{Na}=120 {\rm mS/ cm^2}$, $g_{K}=36 {\rm mS/ cm^2}$, $g_{L}=0.3
{\rm mS/ cm^2}$, $V_{Na}=50$ mV, $V_{K}=-77$ mV, $V_{L}=-54.4$ mV,
and $V_r=-65$ mV.

To obtain the Poincar\'{e} map of Eq.~(\ref{eq:HH1}), we make a
normalization $f_1 t \rightarrow t$, and then Eq.~(\ref{eq:HH1})
can be reduced to the following differential equations:
\begin{subequations}
\begin{eqnarray}
\frac{dV}{dt} &=& F_1({\bf{x}},\theta) \nonumber \\
&=& {\frac {1} {C~f_1}} [-g_{Na}m^3 h (V-V_{Na}) - g_K
n^4 (V-V_K) - g_L (V-V_L) + I_{ext}], \\
\frac{dm}{dt} &=& F_2({\bf{x}},\theta) = {\frac {1} {f_1}}
[\alpha_m (V) (1-m) - \beta_m (V) m], \\
\frac{dh}{dt} &=& F_3({\bf{x}},\theta) = {\frac {1} {f_1}}
[\alpha_h (V) (1-h) - \beta_h (V) h], \\
\frac{dn}{dt} &=& F_4({\bf{x}},\theta) = {\frac {1} {f_1}}
[\alpha_n (V) (1-n) - \beta_n (V) n], \\
\frac {d\theta} {dt} &=& \omega~~{\rm {(mod~1)}},
\end{eqnarray}
\label{eq:HH2}
\end{subequations}
where ${\bf x} [=(x_1,x_2,x_3,x_4)] \equiv (V,m,h,n)$ and $I_{
ext}= I_{dc} + A_1 \sin(2 \pi t) + A_2 \sin(2 \pi \theta)$. The
phase space of the quasiperiodically forced HH oscillator is six
dimensional with coordinates $V$, $m$, $h$, $n$, $\theta$, and
$t$. Since the system is periodic in $\theta$ and $t$, they are
circular coordinates in the phase space. Then, we consider the
surface of section, the $V$-$m$-$h$-$n$-$\theta$ hypersurface at
$t=n$ ($n$: integer). The phase-space trajectory intersects the
surface of section in a sequence of points. This sequence of
points corresponds to a mapping on the five-dimensional
hypersurface. The map can be computed by stroboscopically sampling
the orbit points ${\bf v}_n$ $[\equiv ({\bf x}_n,\theta_n)]$ at
the discrete time $n$ (corresponding to multiples of the first
external driving period $T_1$). We call the transformation ${\bf
v}_n \rightarrow {\bf v}_{n+1}$ the Poincar{\'e} map, and write
${\bf v}_{n+1} = P({\bf v}_n)$.

Numerical integration of Eqs.~(\ref{eq:HH1}) and (\ref{eq:HH2}) is
done using the fourth-order Runge-Kutta method. Dynamical analysis
is performed in both the continuous-time system ({\it i.e.}, flow)
and the discrete-time system ({\it i.e.}, Poincar\'{e} map). For
example, the time series of the membrane potential $V(t)$ and the
phase flow are obtained in the flow. On the other hand, the
Lyapunov exponent \cite{Lexp} and the phase sensitivity exponent
\cite{PD} of an attractor are calculated in the Poincar\'{e} map.
To obtain the Lyapunov exponent of an attractor in the
Poincar\'{e} map, we choose 20 random initial points $\{ (V_i(0),
m_i(0), h_i(0), n_i(0), \theta_i(0)); i=1,\dots,20 \}$ with
uniform probability in the range of $V_i(0) \in (-60,0)$, $m_i(0)
\in (0.1,0.9)$, $h_i(0) \in (0.1,0.2)$, $n_i(0) \in (0.5,0.7)$ and
$\theta_i(0) \in [0,1)$. For each initial point, we get the
Lyapunov exponent, and choose the average value of the 20 Lyapunov
exponents. (The method of obtaining the phase sensitivity exponent
will be explained below.)

\begin{figure}
\includegraphics[width=0.8\columnwidth]{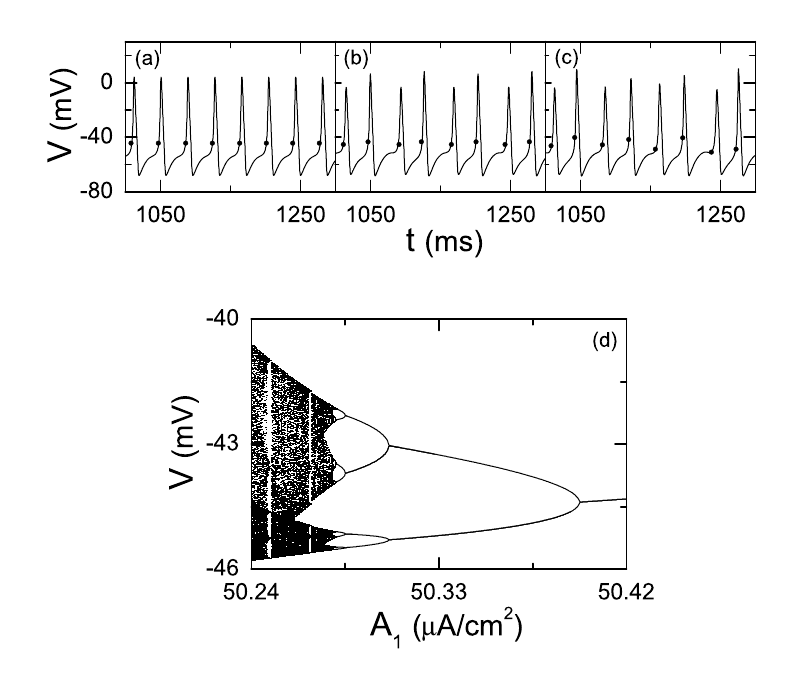}
\caption{Period-doubling route to chaos in the periodically forced
HH oscillator for $I_{dc}=100$ $\mu {\rm A/cm^2}$ and $f_1=26$ Hz
($A_2=0$). Time series of $V(t)$ for (a) $A_1=50.42$ $\mu {\rm
A/cm^2}$, (b) $A_1=50.33$ $\mu {\rm A/cm^2}$, and (c) $A_1=50.24$
$\mu {\rm A/cm^2}$, which correspond to the period-1, period-2,
and chaotic states in the Poincar\'{e} map $P$ (solid circles
represent stroboscopically sampled points in $P$), respectively.
The largest Lyapunov exponent for the chaotic oscillation in (c)
is $\sigma_1 \simeq 0.096$. (d) Bifurcation diagram ({\it i.e.},
plot of $V$ versus $A_1$) in $P$. \label{fig:PD}}
\end{figure}

In the presence of only the dc stimulus (${\it i.e.,}$
$A_1=A_2=0$), a transition from a resting state to a periodic
spiking state occurs for $I_{dc}=I_{dc}^*$ $(\simeq 9.78$ $\mu
{\rm A/cm^2})$ via a subcritical Hopf bifurcation when the resting
state absorbs the unstable limit cycle born via a fold limit cycle
bifurcation for $I_{dc} \simeq 6.26$ $\mu {\rm A/cm^2}$
\cite{HHB,HHB2}. Thus, a self-sustained oscillation (corresponding
to a spiking state) is induced in the HH neuron model for $I_{dc}
> I_{dc}^*$. Here, we set $I_{dc}=100$ $\mu {\rm A/cm^2}$ and
$\omega$ to be the reciprocal of the golden mean [{\it i.e.},
$\omega = (\sqrt{5}-1)/2$], and numerically investigate dynamical
responses of the (self-sustained) HH oscillator under the ac
external stimulus. We first study the case of periodic forcing
({\it i.e.}, $A_2=0$) by varying $A_1$ for $f_1=26$ Hz. Figures
\ref{fig:PD}(a)-\ref{fig:PD}(c) show the time series of $V(t)$ for
$A_1=50.42$, $50.33$, and $50.24$ $\mu {\rm A/cm^2}$,
respectively, and the bifurcation diagram in the Poincar\'{e} map
$P$ is also given in Fig.~\ref{fig:PD}(d); stroboscopically
sampled points in $P$ are represented by solid circles in
Figs.~\ref{fig:PD}(a)-\ref{fig:PD}(c). As $A_1$ is decreased,
successive period-doubling bifurcations occur. For example,
periodic oscillations of $V(t)$ in Figs.~\ref{fig:PD}(a) and
\ref{fig:PD}(b) correspond to period-1 and period-2 states in $P$,
respectively. When $A_1$ passes a threshold $A_1^*$ $(\simeq
50.28$ $\mu {\rm A/cm^2}$) a chaotic transition occurs. Thus, for
$A_1 < A_1^*$ chaotic oscillations with positive Lyapunov
exponents appear, as shown in Fig.~\ref{fig:PD}(c).

\begin{figure}
\includegraphics[width=0.8\columnwidth]{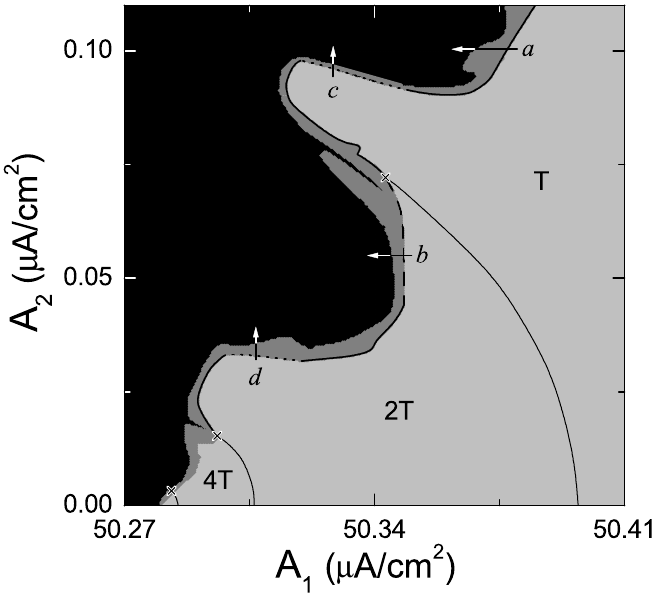}
\caption{State diagram in the $A_1-A_2$ plane for $I_{dc}=100$
$\mu {\rm A/cm^2}$ and $f_1=26$ Hz in the quasiperiodically forced
HH oscillator. Regular, SN, and chaotic regions are shown in light
gray, gray, and black, respectively. In the regular region, the
torus, the doubled torus, and the quadrupled torus exist in the
regions denoted by $T$, $2T$, and $4T$, respectively, and the
solid lines represent torus doubling bifurcation curves with
terminal points denoted by crosses. SN attractors appear via
fractalization, collision with a smooth unstable torus, and
collision with a nonsmooth ring-shaped unstable set when passing
the heavy solid line, the dashed line, and the dotted line,
respectively. \label{fig:SD}}
\end{figure}

From now on, we investigate the effect of quasiperiodic forcing on
the period-doubling route to chaotic oscillation by changing $A_1$
and $A_2$ for $f_1=26$ Hz. Figure \ref{fig:SD} shows a state
diagram in the $A_1-A_2$ plane. Each state is characterized by the
largest (nontrivial) Lyapunov exponent $\sigma_1$, associated with
dynamics of the variable $\bf{x}$ [besides the (trivial) zero
exponent, related to the phase variable $\theta$ of the
quasiperiodic forcing] and the phase sensitivity exponent
$\delta$. The exponent $\delta$ measures the sensitivity of the
variable ${\bf x}$ with respect to the phase $\theta$ of the
quasiperiodic forcing and characterizes the strangeness of an
attractor \cite{PD}. Regular quasiperiodic oscillations occur on
smooth tori. A smooth torus that has a negative largest Lyapunov
exponent ({\it i.e.}, $\sigma_1 <0$) and has no phase sensitivity
({\it i.e.}, $\delta=0$) exists in the region denoted by $T$ and
shown in light gray. When crossing a solid line, the smooth torus
becomes unstable and bifurcates to a smooth doubled torus in the
region represented by $2T$. Smooth quadrupled tori, bifurcated
from doubled tori, also exist in the region denoted by $4T$. On
the other hand, chaotic oscillating states with positive largest
Lyapunov exponents ($\sigma_1>0$) exist in the region shown in
black. Between these regular and chaotic regions, SN oscillating
states that have negative largest Lyapunov exponents
($\sigma_1<0$) and positive phase sensitivity exponents
$(\delta>0)$ exist in the region shown in gray. Due to their high
phase sensitivity, SN oscillating states have a strange fractal
phase space structure. Various dynamical routes to SN oscillations
via gradual fractalization, collision with a smooth unstable
torus, and collision with a nonsmooth ring-shaped unstable set
will be discussed below.

When passing a heavy solid boundary curve in Fig.~\ref{fig:SD}, a
transition from a smooth torus to an SN attractor occurs via
gradual fractalization \cite{NK}. As an example, we study such
transition to SN oscillations along the route $a$ by decreasing
$A_1$ for $A_2=0.1$ $\mu {\rm A/cm^2}$. Figures
\ref{fig:FRAC}(a)-\ref{fig:FRAC}(c) show the time series of $V(t)$
for the quasiperiodic oscillation, the SN oscillation, and the
chaotic oscillation when $A_1=50.41$, $50.374$, and $50.36$ $\mu
{\rm A/cm^2}$, respectively. These regular, SN, and chaotic states
are analyzed in terms of the largest Lyapunov exponent $\sigma_1$
and the phase sensitivity exponent $\delta$ in the Poincar\'{e}
map. Projections of their corresponding attractors onto the
$\theta-V$ plane are shown in
Figs.~\ref{fig:FRAC}(d)-\ref{fig:FRAC}(f). For the regular state,
a smooth torus exists in the $\theta-V$ plane [see
Fig.~\ref{fig:FRAC}(d)]. As $A_1$ is decreased, the smooth torus
becomes more and more wrinkled and transforms to an SN attractor
without apparent mediation of any nearby unstable invariant set
\cite{SNA,PNR}. As an example, see an SN attractor in
Fig.~\ref{fig:FRAC}(e). This kind of gradual fractalization is the
most common route to SN attractors. With further decrease in
$A_1$, such an SN attractor turns into a chaotic attractor, as
shown in Fig.~\ref{fig:FRAC}(f).

A dynamical property of each attractor is characterized in terms
of the largest Lyapunov exponent $\sigma_1$ (measuring the degree
of sensitivity to initial conditions). The Lyapunov-exponent
diagram ({\it i.e.}, plot of $\sigma_1$ vs. $A_1$) is given in
Fig.~\ref{fig:FRAC}(g). When passing a threshold value of $A_1
\simeq 50.377$ $\mu {\rm A/cm^2}$, an SN attractor appears. The
graph of $\sigma_1$ for the SN attractor is shown in black, and
its value is negative as in the case of smooth torus. However, as
$A_1$ passes the chaotic transition point of $A_1 \simeq 50.368$
$\mu {\rm A/cm^2}$, a chaotic attractor with a positive $\sigma_1$
appears. Although SN and chaotic attractors are dynamically
different, both of them have strange geometry. To characterize the
strangeness of an attractor, we investigate the sensitivity of the
attractor with respect to the phase $\theta$ of the external
quasiperiodic forcing \cite{PD}. This phase sensitivity may be
characterized by differentiating $\bf x$ with respect to $\theta$
at a discrete time $t=n$. Using Eq.~(\ref{eq:HH2}), we may obtain
the following governing equation for $\frac {\partial x_i}
{\partial \theta}$ $(i=1,2,3, 4)$,
\begin{equation}
  {\frac {d} {dt}} \left( {\frac {\partial x_i} {\partial \theta}}
  \right) =
  \sum_{j=1}^4 {\frac {\partial F_i} {\partial x_j}}
  \cdot{\frac {\partial x_j}   {\partial \theta}} +
  {\frac {\partial F_i} {\partial \theta}},
\label{eq:PSE}
\end{equation}
where $(x_1,x_2,x_3,x_4) = (V,m,h,n)$ and $F_i$'s $(i=1,2,3, 4)$
are given in Eq.~(\ref{eq:HH2}). Starting from an initial point
$({\bf{x}}(0),\theta(0))$ and an initial value $\partial \bf{x} /
\partial \theta = {\bf 0}$ for $t=0$, we may obtain the derivative
values of $S^{(i)}_n$ $(\equiv
\partial x_i / \partial \theta)$ at all subsequent discrete
time $t=n$ by integrating Eqs.~(\ref{eq:HH2}) and (\ref{eq:PSE}).
One can easily see the boundedness of $S_n^{(i)}$ by looking only
at the maximum
\begin{equation}
\gamma_N^{(i)}({\bf{x}}(0),\theta(0)) = \max_{0 \leq n \leq N}
|S_n^{(i)}({\bf{x}}(0),\theta(0))| \,\,\, (i=1, 2,3, 4).
\label{eq:gFtn}
\end{equation}
We note that $\gamma_N^{(i)}({{\bf x}(0)},\theta(0))$ depends on a
particular trajectory. To obtain a ``representative'' quantity
that is independent of a particular trajectory, we consider an
ensemble of randomly chosen initial points $\{ {\bf
x}(0),\theta(0) \}$, and take the minimum value of
$\gamma_N^{(i)}$ with respect to the initial orbit points
\cite{PD},
\begin{equation}
\Gamma_N^{(i)} = \min_{ \{ {\bf x}(0),\theta(0) \} }
\gamma_N^{(i)}( {\bf x}(0)) \,\,\, (i=1, 2, 3, 4). \label{eq:GFtn}
\end{equation}
Figure \ref{fig:FRAC}(h) shows a phase sensitivity function
$\Gamma_N^{(1)}$, which is obtained in an ensemble containing 20
random initial orbit points $\{ (V_i(0), m_i(0), h_i(0), n_i(0),
\theta_i(0)); i=1,\dots,20 \}$ which are chosen with uniform
probability in the range of $V_i(0) \in (-60,0)$, $m_i(0) \in
(0.1,0.9)$, $h_i(0) \in (0.1,0.2)$, $n_i(0) \in (0.5,0.7)$ and
$\theta_i(0) \in [0,1)$. For the case of the smooth torus in
Fig.~\ref{fig:FRAC}(d), $\Gamma_N^{(1)}$ grows up to the largest
possible value of the derivative $|\partial x_1 /
\partial \theta|$ along a trajectory and remains for all
subsequent time. Thus, $\Gamma_N^{(1)}$ saturates for large $N$
and hence the smooth torus has no phase sensitivity ({\it i.e.},
it has smooth geometry). On the other hand, for the case of the SN
attractor in Fig.~\ref{fig:FRAC}(e), $\Gamma_N^{(i)}$ grows
unboundedly with the same power $\delta$, independently of $i$,
\begin{equation}
\Gamma^{(i)}_N \sim N^\delta. \label{eq:Gamma}
\end{equation}
Here, the value of $\delta \simeq 2.39$ is a quantitative
characteristic of the phase sensitivity of the SN attractor, and
$\delta$ is called the phase sensitivity exponent. For obtaining
satisfactory statistics, we consider 20 ensembles for each $A_1$,
each of which contains 20 randomly chosen initial points and
choose the average value of the 20 phase sensitivity exponents
obtained in the 20 ensembles. Figure \ref{fig:FRAC}(i) shows a
plot of $\delta$ versus $\Delta A_1$ $(=A_1-A_1^*)$. Note that the
value of $\delta$ monotonically increases from zero as $A_1$ is
decreased away from the SN transition point $A_1^*$ $(\simeq
50.377$ $\mu {\rm A/cm^2})$. As a result of this phase
sensitivity, the SN oscillating state has strange fractal geometry
leading to aperiodic complex spikings, as in the case of chaotic
oscillations [{\it e.g.,} see Figs.~\ref{fig:FRAC}(b) and
\ref{fig:FRAC}(c)].

\begin{figure}
\includegraphics[width=0.8\columnwidth]{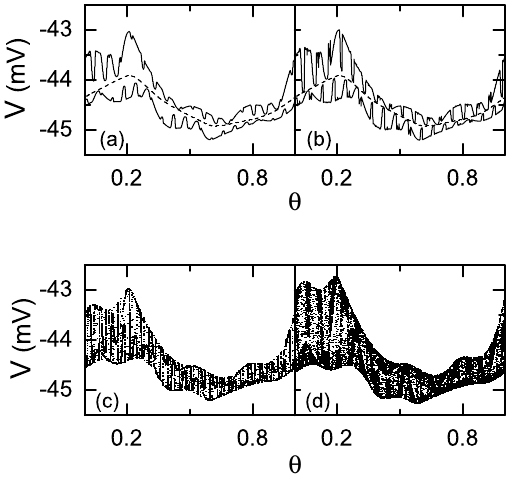}
\caption{Appearance of an SN attractor via phase-dependent
(nonsmooth) collision between a stable two-band torus and its
unstable smooth parent torus along the route $b$ in
Fig.~\ref{fig:SD} for $A_2=0.06$ $\mu {\rm A/cm^2}$. Stable
two-band torus (denoted by a solid curve) and its unstable smooth
torus (represented by a short-dashed curve) for (a) $A_1=50.348$
$\mu {\rm A/cm^2}$ and (b) $A_1=50.347$ $\mu {\rm A/cm^2}$. (c) SN
attractor with $\sigma_1 \simeq -0.015$ and $\delta \simeq 3.77$
for $A_1=50.346$ $\mu {\rm A/cm^2}$. (d) Chaotic attractor with
$\sigma_1 \simeq 0.038$ for $A_1=50.34$ $\mu {\rm A/cm^2}$.
\label{fig:HH}}
\end{figure}

As a dashed boundary curve in Fig.~\ref{fig:SD} is crossed,
another route to SN attractors appears through collision between a
stable smooth doubled torus and its unstable smooth parent torus
\cite{HH}. As an example, we study this transition to SN
oscillations along the route $b$ by decreasing $A_1$ for
$A_2=0.06$ $\mu {\rm A/cm^2}$. Figure \ref{fig:HH} shows a stable
two-band torus (denoted by a solid curve) and an unstable smooth
one-band parent torus (denoted by a short-dashed curve) for
$A_1=50.348$ $\mu {\rm A/cm^2}$. The unstable parent torus is
located in the middle of the two bands of the the stable torus. As
$A_1$ is decreased, the bands of the stable torus becomes more and
more wrinkled, while the unstable torus remains smooth [see
Fig.~\ref{fig:HH}(b)]. When $A_1$ passes a threshold value of $A_1
\simeq 50.3469$ $\mu {\rm A/cm^2}$, the two bands of the stable
torus touch its unstable parent torus at a dense set of $\theta$
values (not at all $\theta$ values). As a result of this
phase-dependent (nonsmooth) collision between the stable doubled
torus and its unstable parent torus, an SN attractor is born, as
shown in Fig.~\ref{fig:HH}(c). This SN attractor, containing the
former bands of the torus as well as the unstable parent torus,
has a positive phase sensitivity exponent (${\it i.e.,}$ $\delta
>0$), inducing the strangeness of the SN attractor. However, its
dynamics is nonchaotic because the largest Lyapunov exponent is
negative (${\it i.e.,}$ $\sigma_1 < 0$). As another threshold
value of $A_1 \simeq 50.344$ $\mu {\rm A/cm^2}$ is passed, the SN
attractor transforms to a chaotic attractor with a positive
largest Lyapunov exponent $\sigma_1$ [see Fig.~\ref{fig:HH}(d)].

\begin{figure}
\includegraphics[width=0.6\columnwidth]{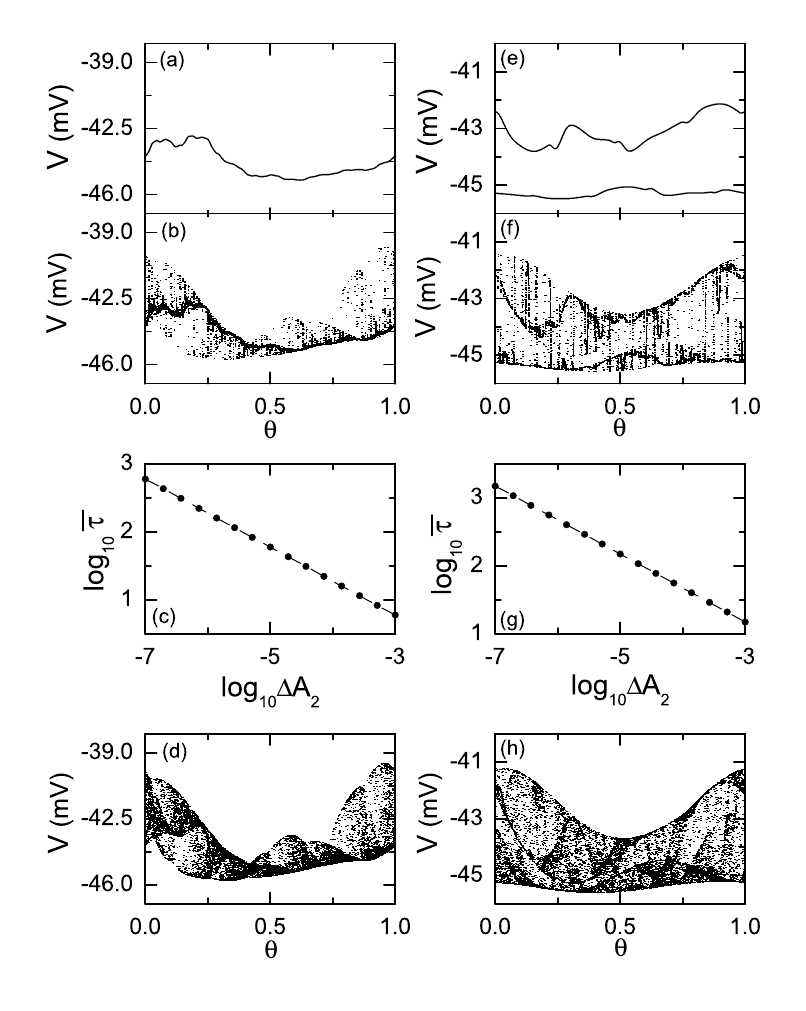}
\caption{Appearance of intermittent SN attractors. (a)-(d)
Appearance of an intermittent SN attractor along the route $c$ in
Fig.~\ref{fig:SD} for $A_1=50.34$ $\mu {\rm A/cm^2}$. (a) Smooth
torus with $\sigma_1 \simeq -0.071$ for $A_2=0.093$ $\mu {\rm
A/cm^2}$. (b) Intermittent SN attractor with $\sigma_1 \simeq
=-0.015$ and $\delta \simeq 3.15$ for $A_2=0.09353$ $\mu {\rm
A/cm^2}$. (c) Plot of ${\rm log_{10}} \overline{\tau}$ vs. ${\rm
log_{10}} \Delta A_2$. The graph is well fitted with a dashed
straight line with slope $\gamma \simeq 0.5$. Here
$\overline{\tau}$ is the average characteristic time between
bursts and $\Delta A_2 = A_2 - A_2^*$ $(A_2^* = 0.093\,35$ $\mu
{\rm A/cm^2})$. For each $\Delta A_2$, $\overline{\tau}$ is
calculated from $10^4$ laminar phases in the Poincar\'{e} map $P$.
(d) Chaotic attractor with $\sigma_1 \simeq 0.083$ for $A_2=0.095$
$\mu {\rm A/cm^2}$. (e)-(h) Band-merging transition from a
two-band torus to an intermittent single-band SN attractor along
the route $d$ in Fig.~\ref{fig:SD} for $A_1=50.3$ $\mu {\rm
A/cm^2}$. (e) Smooth two-band torus with $\sigma_1 \simeq -0.135$
for $A_2=0.03$ $\mu {\rm A/cm^2}$. (f) Intermittent single-band SN
attractor with $\sigma_1 \simeq -0.029$ and $\delta \simeq 2.17$
for $A_2=0.0336$ $\mu {\rm A/cm^2}$. (g) Plot of ${\rm log_{10}}
\overline{\tau}$ vs. ${\rm log_{10}} \Delta A_2$. The graph is
well fitted with a dashed straight line with slope $\gamma \simeq
0.5$. Here $\overline{\tau}$ is the average characteristic time
between bursts and $\Delta A_2 = A_2 - A_2^*$ $(A_2^* = 0.033\,45$
$\mu {\rm A/cm^2}$). For each $\Delta A_2$, $\overline{\tau}$ is
calculated from $10^4$ laminar phases in $P^2$. (h) Chaotic
attractor with $\sigma_1 \simeq 0.079$ for $A_2=0.04$ $\mu {\rm
A/cm^2}$. \label{fig:INT}}
\end{figure}

A main interesting feature of the state diagram in
Fig.~\ref{fig:SD} is the existence of ``tongues'' of quasiperiodic
motion that penetrate into the chaotic region. The first-order
(second-order) tongue lies near the terminal point (denoted by a
cross) of the first-order (second-order) torus-doubling
bifurcation curve. When crossing the upper boundary of the tongue
(denoted by a dotted line), an intermittent SN attractor appears
via phase-dependent collision of a stable torus with a nonsmooth
ring-shaped unstable set \cite{Kim,Kim2}. We first study the
transition to an intermittent SN attractor along the route $c$ in
the first-order tongue by increasing $A_2$ for $A_1=50.34$ $\mu
{\rm A/cm^2}$. Figure \ref{fig:INT}(a) shows a smooth torus for
$A_2=0.093$ $\mu {\rm A/cm^2}$. When passing a threshold value of
$A_2 \simeq 0.093\,35$ $\mu {\rm A/cm^2}$, a sudden transition to
an intermittent SN attractor occurs, as shown in
Fig.~\ref{fig:INT}(b) for $A_2=0.093\,53$. Due to high phase
sensitivity, this SN attractor with $\delta \simeq 3.15$ has a
strange fractal structure, while its dynamics is nonchaotic
because of a negative largest Lyapunov exponent ($\sigma_1 \simeq
-0.015$). A typical trajectory on the intermittent SN attractor
spends a long stretch of time in the vicinity of the former torus,
then it bursts out from this region and traces out a much larger
fraction of the state space, and so on. To characterize the
intermittent bursting, we use a small quantity $d^*$ for the
threshold value of the magnitude of the deviation from the former
torus. When the deviation is smaller (larger) than $d^*$, the
intermittent attractor is in the laminar (bursting) phase. For
each $A_2$, we follow a long trajectory until $10^4$ laminar
phases are obtained in the Poincar\'{e} map $P$ and get the
average of characteristic time $\tau$ between bursts. As shown in
Fig.~\ref{fig:INT}(c), the average value of $\tau$ exhibits a
power-law scaling behavior,
\begin{equation}
\overline{\tau} \sim \Delta A_2^{-\gamma},~~\gamma \simeq 0.5,
\end{equation}
where the overbar represents time averaging and $\Delta A_2^* =
A_2 - A_2^*$ $(A_2^*=0.093\,35)$. The scaling exponent $\gamma$
seems to be the same as that for the case of the quasiperiodically
forced map \cite{PMR}. As $A_2$ passes another threshold value of
$A_2 \simeq 0.0936$ $\mu {\rm A/cm^2}$, the SN attractor
transforms to a chaotic attractor because the largest Lyapunov
exponent $\sigma_1$ becomes positive, as shown in
Fig.~\ref{fig:INT}(d). Furthermore, using the rational
approximation, the mechanism for the intermittent route to SN
attractors will be investigated below. Thus, a smooth torus is
found to transform to an intermittent SN attractor via
phase-dependent collision with a nonsmooth ring-shaped unstable
set.

We also study another intermittent route to SN attractors along
the route $d$ in the second-order tongue by increasing $A_2$ for
$A_1=50.3$ $\mu {\rm A/cm^2}$. Figure \ref{fig:INT}(e) shows a
smooth two-band torus for $A_2=0.03$ $\mu {\rm A/cm^2}$. As $A_2$
passes a threshold value of $A_2 \simeq 0.033\,45$ $\mu {\rm
A/cm^2}$, a band-merging transition from a smooth doubled torus to
a single-band SN attractor occurs \cite{Kim2}. Thus, an
intermittent single-band SN attractor appears [{\it e.g.,} see the
intermittent SN attractor with $\sigma_1 \simeq -0.029$ and
$\delta \simeq 2.17$ in Fig.~\ref{fig:INT}(f) for $A_2=0.0336$]. A
typical trajectory of the second iterate of the Poincar\'{e} map
$P$ ({\it i.e.} $P^2$) spends a long stretch of time near one of
the two former attractors ({\it i.e.,} smooth tori), then it
bursts out of this region and comes close to the same or other
former torus where it remains again for some time interval, and so
on. As in the above case of intermittent route to SN attractors,
we also obtain the $10^4$ laminar phases from a long trajectory in
$P^2$, and get the average of characteristic time $\tau$ between
bursts. As shown in Fig.~\ref{fig:INT}(g), the average
characteristic time shows a power-law scaling behavior,
\begin{equation}
\overline{\tau} \sim \Delta A_2^{-\gamma},~~\gamma \simeq 0.5,
\end{equation}
where $\Delta A_2 = A_2 - A_2^*$ $(A_2^*=0.033\,45)$. The scaling
exponent $\gamma$ seems to be the same as that for the case of the
intermittent route to SN attractors occurring near the first-order
tongue. Since the dynamical mechanism for the appearance of
intermittent SN attractors near the first-order and second-order
tongues are the same ({\it i.e.,} an intermittent SN attractor
appears via a phase-dependent collision between a smooth torus and
a nonsmooth ring-shaped unstable set), the intermittent SN
attractors for both cases seem to exhibit the same scaling
behaviors. As $A_2$ is further increased and passes another
threshold value of $A_2 \simeq 0.0352$ $\mu {\rm A/cm^2}$, the SN
attractor turns into a chaotic attractor with a positive largest
Lyapunov exponent $\sigma_1$, as shown in Fig.~\ref{fig:INT}(h).

\begin{figure}
\includegraphics[width=0.6\columnwidth]{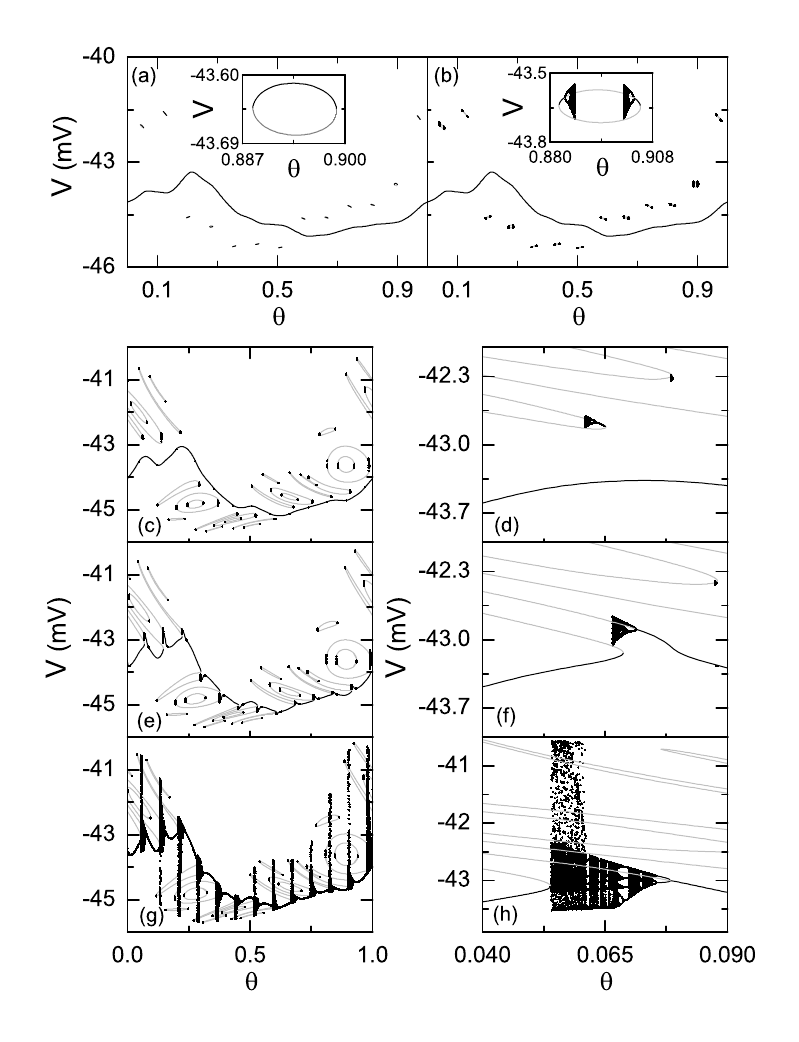}
\caption{Intermittent transition to an SN attractor in the
rational approximation. (a)-(b) Birth and evolution of a nonsmooth
ring-shaped unstable set for $A_2=0.08$ $\mu {\rm A/cm^2}$ in the
rational approximation of level $k=7$. (a) Smooth torus (denoted
by a black curve) and ring-shaped unstable set for $A_1=50.3432$
$\mu {\rm A/cm^2}$. Each ring is composed of the attracting part
(shown in black) and the unstable part (shown in gray). A
magnified view of a ring is given in the inset. (b) Smooth torus
(represented by a black curve) and an evolved ring-shaped unstable
set for $A_1=50.343$ $\mu {\rm A/cm^2}$. As $A_1$ is decreased,
the unstable part (shown in gray) in each ring becomes dominant.
(c)-(h) Mechanism for the intermittent transition to an SN
attractor along the route $c$ in Fig.~\ref{fig:SD} for $A_1=50.34$
$\mu {\rm A/cm^2}$. (c) Smooth torus and a ring-shaped unstable
set for $A_2=0.089$ $\mu {\rm A/cm^2}$ [a magnified view near
$\theta=0.065$ is given in (d)]. New rings appear inside or
outside the old rings. (e) Nonsmooth attractor born via a
phase-dependent (nonsmooth) collision between the smooth torus and
a ring-shaped unstable set for $A_2=0.0904$ $\mu {\rm A/cm^2}$
[(f) is a magnified view near $\theta=0.065$]. (g) Intermittent SN
attractor with $F_7$ gaps (born via an interior crisis) for
$A_2=0.0915$ $\mu {\rm A/cm^2}$. (h) A magnified gap near
$\theta=0.065$. \label{fig:RA}}
\end{figure}

The dynamical mechanisms for the appearance of intermittent SN
attractors near the tongues are the same, irrespectively of the
tongue order. Here, we consider the case of the main first-order
tongue, and using the rational approximation to the quasiperiodic
forcing, we search for an unstable orbit that causes the
intermittent transition via collision with the smooth torus for
$A_2=0.08$ $\mu {\rm A/cm^2}$. For the inverse golden mean $\omega
[=(\sqrt{5}-1)/2]$, its rational approximants are given by the
ratios of the Fibonacci numbers, $\omega_k = F_{k-1} / F_k$, where
the sequence of $\{ F_k \}$ satisfies $F_{k+1} = F_k + F_{k-1}$
with $F_0 = 0$ and $F_1 = 1$. Instead of the quasiperiodically
forced system with $\omega$, periodically forced systems with
$\omega_k$ are studied in the rational approximation. As an
example, we consider the rational approximation of level $k=7$.
The rational approximation to the smooth torus (denoted by a black
curve), composed of stable orbits with period $F_7(=13)$, is shown
in Fig.~\ref{fig:RA}(a) for $A_1=50.3432$ $\mu {\rm A/cm^2}$. We
note that a ring-shaped unstable set, composed of $F_7$ small
rings, lies near the smooth torus. At first, each ring is composed
of the stable (shown in black) and unstable (shown in gray) orbits
with period $F_7$ [see the inset in Fig.~\ref{fig:RA}(a)].
However, as $A_1$ is changed such rings make evolution, as shown
in Fig.~\ref{fig:RA}(b) for $A_1=50.343$ $\mu {\rm A/cm^2}$. For
fixed values of $A_1$ and $A_2$, the phase $\theta$ may be
regarded as a ``bifurcation parameter.'' As $\theta$ is varied, a
chaotic attractor appears via an infinite sequence of
period-doubling bifurcations of stable periodic orbits in each
ring, and then it disappears via a boundary crisis when it
collides with the unstable $F_7$-periodic orbit [see the inset in
Fig.~\ref{fig:RA}(b)]. Thus, the attracting part (shown in black)
of each ring is composed of the union of the originally stable
$F_7$-periodic attractor, the higher $2^n F_7$-periodic
$(n=1,2,\dots)$ and chaotic attractors born through the
period-doubling cascade. On the other hand, the unstable part
(shown in gray) of each ring consists of the union of the
originally unstable $F_7$-periodic orbit [{\it i.e.,} the lower
gray line in the inset in Fig.~\ref{fig:RA}(b)] and the
destabilized $F_7$-periodic orbit [{\it i.e.,} the upper gray line
in the inset in Fig.~\ref{fig:RA}(b)] via a period-doubling
bifurcation. As the parameters, $A_1$ and $A_2$, are further
changed, both the size and shape of the rings change, and
eventually each ring is composed of a large unstable part (shown
in gray) and a small attracting part (shown in black), as shown in
Fig.~\ref{fig:RA}(c) for $A_1=50.34$ $\mu {\rm A/cm^2}$ and
$A_2=0.089$ $\mu {\rm A/cm^2}$ [a magnified view is given in
Fig.~\ref{fig:RA}(d)]. We also note that new rings appear inside
or outside the ``old'' rings.

Finally, in terms of the rational approximation of level 7, we
explain the mechanism for the intermittent transition occurring in
the first-order tongue for $A_1=50.34$ $\mu {\rm A/cm^2}$ [see
Figs.~\ref{fig:INT}(a)-\ref{fig:INT}(b)]. As we approach the
border of the intermittent transition in the state diagram, the
ring-shaped unstable set comes closer to the smooth torus, as
shown in Fig.~\ref{fig:RA}(c) for $A_2=0.089$ $\mu {\rm A/cm^2}$
[see a magnified view in Fig.~\ref{fig:RA}(d)]. As $A_2$ passes a
threshold value of $A_2 \simeq 0.090\,305$ $\mu {\rm A/cm^2}$, a
phase-dependent (nonsmooth) collision occurs between the smooth
torus and the unstable part (shown in gray) of the nonsmooth
ring-shaped unstable set. Then, the new attractor of the system
contains the attracting part (shown in black) of the ring-shaped
unstable set and becomes nonsmooth, which is shown in
Fig.~\ref{fig:RA}(e) for $A_2=0.0904$ $\mu {\rm A/cm^2}$ [see a
magnified view in Fig.~\ref{fig:RA}(f)]. As $A_2$ is further
increased, the chaotic component in the rational approximation to
the attractor increases, and eventually for $A_2 \simeq
0.091\,202$ $\mu {\rm A/cm^2}$, it becomes suddenly widened via an
interior crisis when it collides with the nearest ring [{\it
e.g.,} see Fig.~\ref{fig:RA}(g)]. Then, $F_7(=13)$ ``gaps,'' where
no attractors with period $F_7$ exist, are formed. A magnified gap
is shown in Fig.~\ref{fig:RA}(h). We note that this gap is filled
by intermittent chaotic attractors. Thus, the rational
approximation to the whole attractor consists of the union of the
periodic and chaotic components. For this case, the periodic
component is dominant, and hence the average largest Lyapunov
exponent $(\langle \sigma_1 \rangle \simeq -0.037)$ becomes
negative, where $\langle \cdots \rangle$ denotes the average over
the whole $\theta$. Hence, the rational approximation to the
attractor becomes nonchaotic. We note that the 7th rational
approximation to the attractor in Fig.~\ref{fig:RA}(g) resembles
the (original) intermittent SN attractor in Fig.~\ref{fig:INT}(b),
although the level of the rational approximation is low. In this
way, the intermittent transition to an SN attractor occurs through
two steps in the rational approximation: the phase-dependent
(nonsmooth) collision and the interior crisis.

\section{Summary}
\label{sec:SUM} We have numerically studied dynamical responses of
the quasiperiodically forced HH neural oscillator and compared
them with those for the periodically forced case. For the case of
periodic forcing, a transition from a periodic to a chaotic
oscillation has been found to occur via period doublings in both
numerical and experimental works. Effect of the quasiperiodic
forcing on this period-doubling route to chaotic oscillation has
been investigated. In contrast to the case of periodic forcing,
new type of SN oscillating states have been found to exist between
the regular and chaotic oscillating states as intermediate ones.
Due to their strange geometry, these SN oscillations lead to the
occurrence of aperiodic complex spikings, as in the case of
chaotic oscillations. Hence, SN oscillating states might be a
dynamical origin for the complex spikings which are usually
observed in cortical neurons. Various routes to SN oscillations
via fractalization, collision with a smooth unstable torus, and
collision with a nonsmooth ring-shaped unstable set have been
identified, as in the quasiperiodically forced logistic map
\cite{SNA}. These SN responses are also found to occur in other
neurons exhibiting period-doubling route to chaos ({\it e.g.,} the
Morris-Lecar neuron and the FitzHugh-Nagumo neuron) under
quasiperiodic stimulus \cite{Kim5}. Finally, we suggest an
experiment on the quasiperiodically forced squid giant axon and
expect that SN spikings to be observed. However, the real
biological environment is a noisy one. Hence, it is necessary to
further investigate the effect of noise on the SN response for
real experiment. This type of investigation is beyond the scope of
the the present paper, and hence it is left as a future work.

\begin{acknowledgments}
This work was supported by the Research Grant from the Kangwon
National University. S.-Y. Kim thanks Prof. Yakovlev for
hospitality.
\end{acknowledgments}

\newpage

\begin{figure}
\includegraphics[width=0.8\columnwidth]{fig3}
\caption{Appearance of an SN attractor via fractalization along
the route $a$ in Fig.~\ref{fig:SD} for $A_2=0.1$ $\mu {\rm
A/cm^2}$. Time series of $V(t)$ for (a) the quasiperiodic spiking
state ($A_1=50.41$ $\mu {\rm A/cm^2}$), (b) the SN spiking state
($A_1=50.374$ $\mu {\rm A/cm^2}$), and (c) the chaotic spiking
state ($A_1=50.36$ $\mu {\rm A/cm^2}$). Projections of attractors
onto the $\theta-V$ plane in the Poincar\'{e} map are shown for
(d) the smooth torus [corresponding to (a)], (e) the SN attractor
[corresponding to (b)], and (f) the chaotic attractor
[corresponding to (c)]. (g) Lyapunov-exponent diagram ({\it i.e.},
plot of $\sigma_1$ vs. $A_1$); $\sigma_1$ for the SN attractor is
shown in black. (h) Phase sensitivity functions $\Gamma_N^{(1)}$
are shown for the smooth torus ($T$) [(d)] and the SN attractor
(SNA) [(e)]. For the case of the SN attractor, the graph is well
fitted with a dashed straight line with slope $\delta \simeq
2.39$. (i) Plot of the phase sensitivity exponent $\delta$ versus
$\Delta A_1$ $(=A_1-A_1^*)$ for the SN attractor; $A_1^* \simeq
50.377$. \label{fig:FRAC}}
\end{figure}

\begin{thebibliography}{}
\bibitem{Chicken} M. R. Guevara, L. Glass, and A. Shrier, Science
{\bf 214}, 1350 (1981); L. Glass, M. R. Guevara, A. Shrier, and R.
Perez, Physica D {\bf 7}, 89 (1983).
\bibitem{Squid} K. Aihara, T. Numajiri, G. Matsumoto, and M.
Kotani, Phys. Lett. A {\bf 116}, 313 (1986); N. Takahashi, Y.
Hanyu, T. Musha, R. Kubo, and G. Matsumoto, Physica D {\bf 43},
318 (1990); D. T. Kaplan, J. R. Clay, T. Manning, L. Glass, M. R.
Guevara, and A. Shrier, Phys. Rev. Lett. {\bf 76}, 4074 (1996).
\bibitem{Aih} K. Aihara, Scholarpedia 3(5): 1786 (2008); see also
references therein.
\bibitem{Glass} L. Glass and M. C. Mackey, {\it From Clocks to Chaos}
(Princeton University Press, Princeton, 1988).
\bibitem{Stoop} R. Stoop, K. Schindler, and L.A. Bunimovich,
Neurosci. Res. {\bf 36}, 81 (2000); Nonlinearity {\bf 13}, 1515
(2000).
\bibitem{QO} M. Ding and J. A. S. Kelso, Int. J.
Bifurcation Chaos Appl. Sci. Eng. {\bf 2}, 295 (1992).
\bibitem{SNA} U. Feudel, S. Kuznetsov, and A. Pikovsky, {\it
Strange Nonchaotic Attractors} (World Scientific, Singapore,
2006).
\bibitem{PNR} A. Prasad, S. S. Negi, and R. Ramaswamy, Int. J.
Bifuraction Chaos Appl. Sci. Eng. {\bf 11}, 291 (2001); U. Feudel,
C. Grebogi, and E. Ott, Phys. Rep. {\bf 290}, 11 (1997).
\bibitem{Greb} C. Grebogi, E. Ott, S. Pelikan, and J. A. Yorke, Physica D
{\bf 13}, 261 (1984).
\bibitem{NK} K. Kaneko, Prog. Theor. Phys. {\bf 72}, 202 (1984);
T. Nishikawa and K. Kaneko, Phys. Rev. E {\bf 54}, 6114 (1996).
\bibitem{HH} J. F. Heagy and S. M. Hammel, Physica D {\bf 70}, 140
(1994).
\bibitem{PD} A. S. Pikovsky and U. Feudel, Chaos {\bf 5}, 253
(1995).
\bibitem{PMR} A. Prasad, V. Mehra, and R. Ramaswamy, Phys.\ Rev.\
Lett.\ {\bf 79}, 4127 (1997).
\bibitem{Kim} S.-Y. Kim, W. Lim, and E. Ott, Phys.\ Rev.\ E
{\bf 67}, 056203 (2003); S.-Y. Kim and W. Lim, J. Phys. A {\bf
37}, 6477 (2004).
\bibitem{Kim2} W. Lim and S.-Y. Kim, Phys. Lett. A {\bf 335}, 383
(2005).
\bibitem{Kim3} W. Lim and S.-Y. Kim, Phys. Lett. A {\bf 355}, 331
(2006); S.-Y. Kim and W. Lim, {\it ibid.} {\bf 334}, 160 (2005).
\bibitem{Kim4} J.-W. Kim, S.-Y. Kim, B. Hunt, and E. Ott, Phys.
Rev. E {\bf 67}, 036211 (2003).
\bibitem{Exp} W. L. Ditto, M. L. Spano, H. T. Savage, S. N.
Rauseo, J. Heagy, and E. Ott, Phys. Rev. Lett. {\bf 65}, 533
(1990); W. X. Ding, H. Deutsch, A. Dinklage, and C. Wilke, Phys.
Rev. E {\bf 55}, 3769 (1997); B. P. Bezruchko, A. P. Kuznetsov,
and Y. P. Seleznev, Phys. Rev. E {\bf 62}, 7828 (2000); K.
Thamilmaran, D. V. Senthikumar, A. Venkatesan, and M. Lakshmanan,
Phys. Rev. E {\bf 74}, 036205 (2006).
\bibitem{HH1} A. L. Hodgkin and A. F. Huxley, J. Physiol.\
(London) {\bf 117}, 500 (1952).
\bibitem{HH2} K. Aihara, G. Matsumoto, and Y. Ikegaya, J. Theor.\
Biol.\ {\bf 109}, 249 (1984).
\bibitem{HH3} D. Hansel, G. Mato, and C. Meunier, Europhys.\
Lett.\ {\bf 23}, 367 (1993).
\bibitem{Lexp} A. J. Lichtenberg and M. A. Lieberman, {\it Regular
and Stochastic Motion} (Springer-Verlag, New York, 1983), p. 283.
\bibitem{HHB} E. M. Izhikevich,  Int. J.
Bifurcation Chaos Appl. Sci. Eng. {\bf 10}, 1171 (2000).
\bibitem{HHB2} S.-G. Lee, A. Neiman, and S. Kim, Phys.\ Rev.\ E
{\bf 57}, 3292 (1998).
\bibitem{Kim5} W. Lim, S.-Y. Kim, and Y. Kim (unpublished).

\end{thebibliography}
\end{document}